\documentclass[]{spie}  

\usepackage{amsmath,amsfonts,amssymb}
\usepackage{graphicx}
\usepackage[colorlinks=true, allcolors=blue]{hyperref}

\title{Development of RFSoC-based direct sampling highly multiplexed microwave SQUID readout for future CMB and submillimeter surveys }

\author[a]{Chao Liu}
\author[a,b]{Zeeshan Ahmed}
\author[a,b]{Shawn W. Henderson}
\author[a]{Ryan Herbst}
\author[a]{Larry Ruckman}
\author[b,c]{Thomas Satterthwaite}

\affil[a]{SLAC National Accelerator Laboratory, Menlo Park, CA 94025, USA}
\affil[b]{Kavli Institute for Particle Astrophysics and Cosmology, Stanford, CA 94305, USA}
\affil[c]{Department of Physics, Stanford University, Stanford, CA 94305, USA}

\authorinfo{Correspondence author: Chao Liu. E-mail: chaoliu@slac.stanford.edu}

\begin{document} 
\maketitle
\begin{abstract}
The SLAC Microresonator Radio Frequency (SMuRF) electronics is being deployed as the readout for the Cosmic Microwave Background (CMB) telescopes of the Simons Observatory (SO).  A Radio Frequency System-on-Chip (RFSoC) based readout of microwave frequency resonator based cryogenic sensors is under development at SLAC as an upgrade path for SMuRF with simplified RF hardware, a more compact footprint, and lower total power consumption. The high-speed integrated data converters and digital data path in RFSoC enable direct RF sampling without analog up and down conversion for RF frequencies up to 6 GHz. A comprehensive optimization and characterization study has been performed for direct RF sampling for microwave SQUID multiplexers, which covers noise level, RF dynamic range, and linearity using a prototype implementation. The SMuRF firmware, including the implementation of closed-loop tone tracking, has been ported to the RFSoC platform and interfaced with the quadrature mixers for digital up and down conversion in the data converter data path to realize a full microwave SQUID multiplexer readout. In this paper, a selection of the performance characterization results of direct RF sampling for microwave SQUID multiplexer readout will be summarized and compared with science-driven requirements. Preliminary results demonstrating the read out of cryogenic sensors using the prototype system will also be presented here.  We anticipate our new RFSoC-based SMuRF system will be an enabling readout for on-going and future experiments in astronomy and cosmology, which rely on large arrays of cryogenic sensors to achieve their science goals. 
\end{abstract}


\section{INTRODUCTION}
\label{sec:intro}  

The SLAC microresonator RF (SMuRF) electronics has enabled the readout of high density superconducting detector arrays for a range of cosmic microwave background (CMB) experiments \cite{cukierman2020microwave,yu2023slac}. The instrumentation technology for far-infrared (FIR) to millimeter-wave detection and imaging progresses rapidly in fabricating ever higher channel count superconducting detector arrays which in turn must be read out using increasingly higher density multiplexing techniques.\cite{ost2018origins, karkare2022snowmass}. Many advanced sensor readout architectures implement sensor multiplexing in the frequency domain by coupling the superconducting sensors’ signals to combs of closely spaced superconducting resonators. One such technology is microwave superconducting quantum interference device (SQUID) multiplexing \cite{irwin2004microwave}, for which SLAC has developed the custom SMuRF electronics \cite{yu2023slac}. Microwave SQUID multiplexers developed at NIST \cite{dober2021microwave} read out using the SMuRF electronics have achieved multiplexing factors of O(1000), reading out superconducting transition edge sensor arrays optimized for CMB imaging \cite{dutcher2024simons}. New technologies that have been developed since the custom SMuRF system’s design have the potential to enable significant reductions in the size, weight and power (SWaP) of a future SMuRF-like system, especially due to the fact that SMuRF was implemented using custom boards using discrete data converters and FPGA. With future science goals and strategies pushing towards arrays of even higher sensitivities and densities, readout solutions must be continuously updated to keep pace. To this end, since the release of RFSoC devices from AMD Xilinx, our team started to investigate the feasibility of using the RFSoC as a fully digital upgrade path for the SMuRF electronics. 

The integrated data converters demonstrated high performance for both first Nyquist zone  \cite{liu2021characterizing} and higher order Nyquist zones without analogue mixers \cite{liu2022development,liu2023higher,liu2023evaluating}, which is known as direct RF sampling. Through the exercises of characterizing the performance of the data converters, the optimum configurations for high density readout have been determined. The full SMuRF firmware has been ported to the integrated programming logic (PL) of RFSoC and the SMuRF software has been adapted to be interfaced to the RFSoC for control and data transfer purposes. This new readout SMuRF-RFSoC platform has been implemented with direct RF sampling techniques, eliminating the costly and complex custom RF front-end of the legacy SMuRF system. The simplified digital mixing architecture enabled by the RFSoC offers further advantages in SWaP, potentially making it attractive for balloon and space experiments \cite{lowe2020balloon, bradley2021advancements}. The focus of SMuRF-RFSoC platform prototype development is on astronomy and cosmology applications, but the platform offers high flexibility to be adapted for other physics experiments, from bench-top small scale device testing to large scale readout and control platforms.  Potential applications include axion searches using microwave cavities, dark matter detection with quantum sensors, and even RF control for linear accelerators \cite{liu2024direct}.  

The evaluation and development has been performed with both commercial AMD Xilinx evaluation boards and a custom board designed for space applications \cite{henderson2022advanced}. Our development has focused on the commercial evaluation board to maximize accessibility and flexibility. Regardless of which RFSoC device or platform is used, the software architecture and firmware modules required for the targeted applications have significant commonalities. One goal of this development effort has been to build a library of re-configurable firmware and software modules to enable rapid prototyping targetting diverse applications. In this paper, the system design is discussed, highlighting some of the most critical design trade-offs in the platform.  We also present an initial standalone RF performance evaluation and first cryostat test results using an early prototype SMuRF-RFSoC implementation. A range of performance characterization and optimization work is still in progress with more detailed results to be published in future publications.
  \begin{figure} [ht]
   \begin{center}
   \begin{tabular}{c} 
   \includegraphics[height=7.3cm]{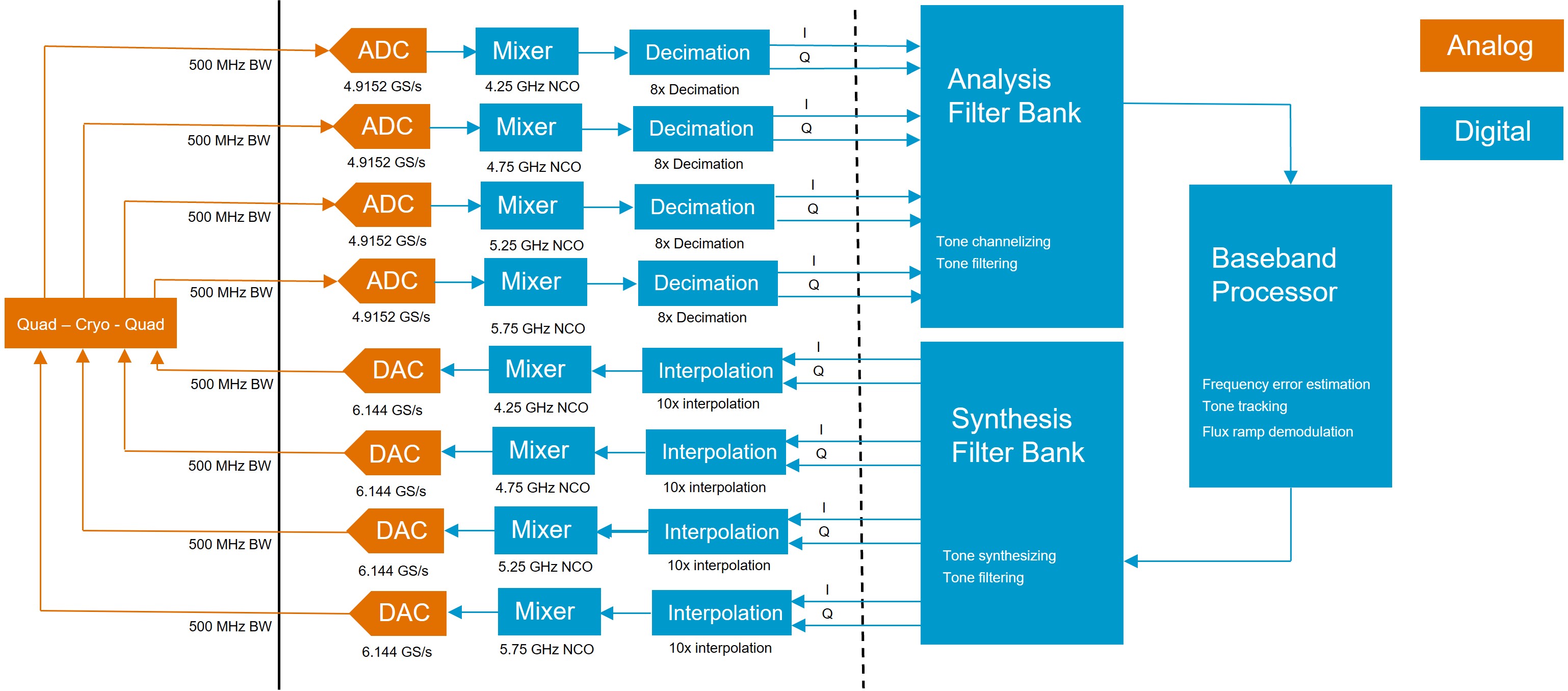}
   \end{tabular}
   \end{center}
   \caption
   { \label{fig:f1} 
The block diagram for the SMuRF-RFSoC system.The solid black line divides the RFSoC and other components of the readout system. The right side of the solid line shows the data-path configurations for both ADCs and DACs and the feedback data processing firmware blocks. All the blocks on the right side of the solid line are implemented digitally within the RFSoC. For the legacy SMuRF electronics, all of the analog mixers, data converters, and FPGA are implemented on multiple custom boards.}
   \end{figure} 

\section{System Design}

The architecture of this prototype SMuRF-RFSoC implementation is shown in Figure \ref{fig:f1}. We have developed the firmware and software to be compatible for multiple commercial evaluation boards, including the AMD ZCU208 and ZCU216 for general purpose evaluation and the smaller RFSoC 4x2 kit for academic evaluation and small scale prototyping \cite{RFSoC4x2}. In this proceedings, the ZCU208 evaluation board with a Gen 3 RFSoC device, the XCZU48DR-2FSVG1517E5184, is used as the prototyping hardware platform. The device integrates 8 ADCs and 8 DACs sampling up to 7 giga sample per second (GSPS) when the integrated digital mixers are enabled. The DACs and ADCs are configured to 6.144 GSPS and 4.9512 GSPS respectively. The SMuRF firmware is built to cover 4 GHz of bandwidth using eight separate contiguous 500 MHz bands. This prototype version of SMuRF-RFSoC is designed to have two readout blocks, each of which covers 4 to 6 GHz using four 500 MHz bands apiece. Figure \ref{fig:f1} shows a block diagram of the implementation of one 4 to 6 GHz block of the SMuRF-RFSoC readout. The input RF signal, typically from a cryostat, is split into four filtered 500 MHz bands by a quadruplexer. The signals are digitized by the ADCs and down-converted at the corresponding centre frequency of each of the four bands. The base-band digital data streams are in in-phase (I) and quadrature (Q) format and decimated by a factor of 8. The I and Q components are channelized in the frequency domain by the analysis filter bank. The sub-bands are processed for multiple purposes, such as flux ramp demodulation, frequency error estimation, and tone tracking, in a time interleaved manner. The baseband processor presents an updated frequency look-up-table and the synthesis filter bank generates the time domain data streams in I and Q format. The I and Q pairs are interpolated by a factor of 10 and then up-converted at the corresponding centre frequencies and loaded to the DACs for generating output RF signals. The four DAC output signals are then filtered and combined into a single output RF signal by a quadruplexer. The RF signal is then injected back into the cryostat for continuous measurement.

\begin{figure} [ht]
   \begin{center}
   \begin{tabular}{c} 
   \includegraphics[height=9cm]{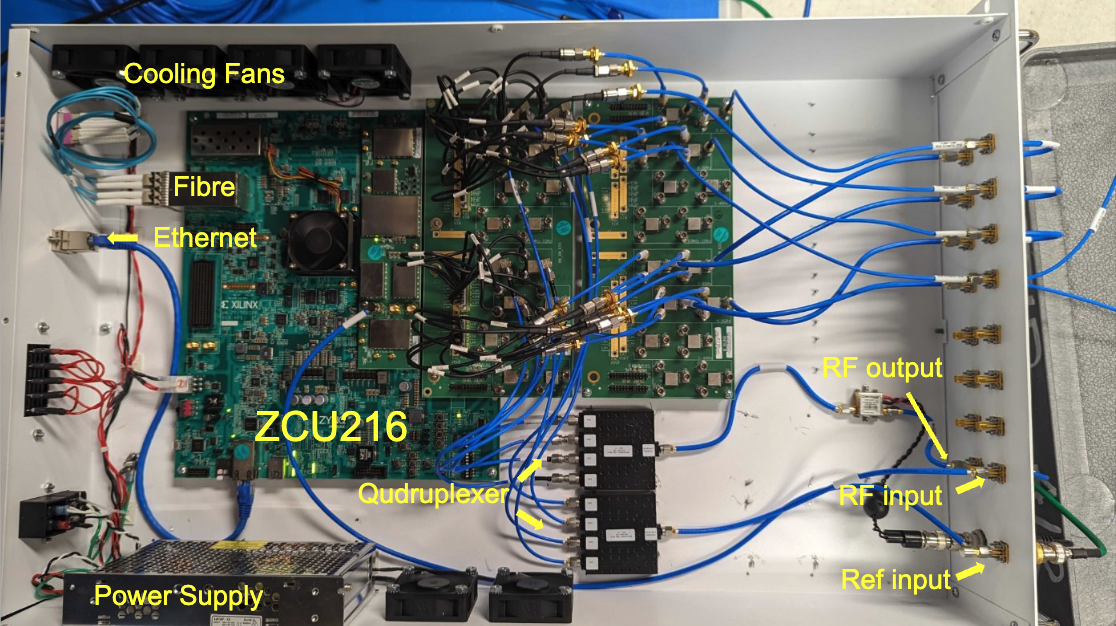}
   \end{tabular}
   \end{center}
   \caption
   { \label{fig:f2} 
The 3U rack mount enclosure designed for SMuRF-RFSoC. The RF inputs and outputs from the ZCU208 evaluation board are routed to a commercial Xilinx RF-FMC daughter card via RF-FPGA Mezzanine Card (FMC) connectors on the ZCU208 board. The RF signals are then converted from differential to single-ended or vice versa using cables connected to balun circuits on a commercial RF-FMC daughter board.}
   \end{figure}

Figure \ref{fig:f2} shows a custom 3U rack mount enclosure designed for the SMuRF-RFSoC. Two instances of readout over 4 to 6 GHz are implemented in the RFSoC firmware with all of the ZCU208's 8 ADCs and 8 DACs used in this implementation. The first readout instance has the full combining and dividing circuit as described above and a high linearity low noise amplifier (LNA) has been added on the DAC side. The ZX60-83LN12+ LNA from Mini-Circuits was selected for this purpose due to its high third order intercept point (IP3) in order to minimize inter-modulation (IM) distortion products. In Figure \ref{fig:f2}, the second readout instance has all four input and four output bands directed routed to the front panel of the enclosure for testing purposes. The fibre connections for data links and timing and Ethernet for remote monitoring and reprogramming are routed to the back panel of the enclosure. The SMuRF-RFSoC is synchronized to the low frequency legacy SMuRF modules and external RF instrumentation via an external 10 MHz clock from master clock source via a reference input. The ZCU208 and cooling fans are driven by a 150W power supply unit, which has a 12V output.

\section{RF Performance Characterization}

After installation in the enclosure, the RF performance of the SMuRF-RFSoC platform was characterized with the support of existing SMuRF software packages. For the RF performance characterization, the first instance of the system is configured in loop-back mode with the combined RF output directly connected to the RF input. For most superconducting sensor readout technologies, the RF dynamic range is one of the most crucial performance characteristics.  We therefore characterized the RF dynamic range of the system by generating and reading back single RF tones as a function of frequency in loop-back mode across the full 4-6 GHz bandwidth.  The RF dynamic range is measured as a function of tone frequency in each 500~MHz band using the SMuRF software to configure the system to generate tones and capture resulting data stream.  After downconversion, the base-band signal is channelized by the analysis filter bank and captured for further processing using the SMuRF software. 

\begin{figure} [ht]
   \begin{center}
   \begin{tabular}{c} 
   \includegraphics[height=7cm]{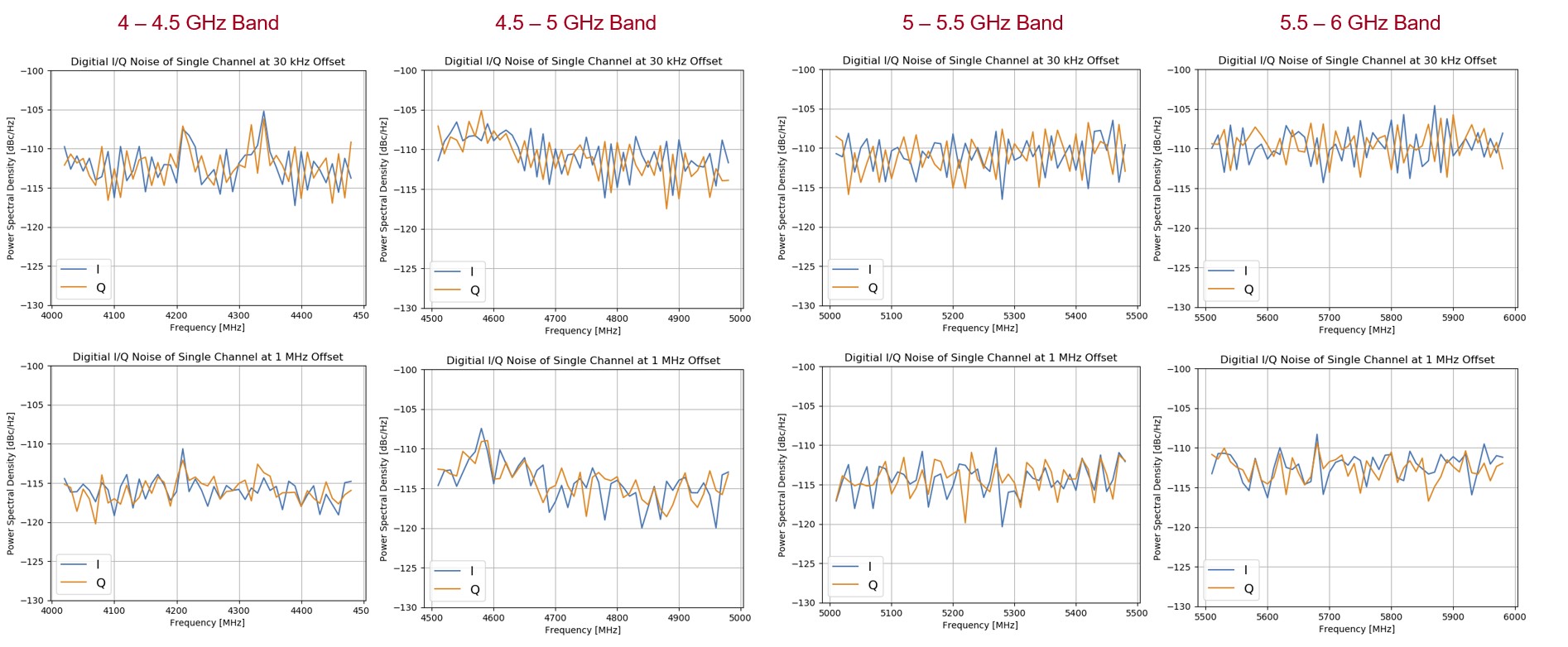}
   \end{tabular}
   \end{center}
   \caption
   { \label{fig:f3} 
Measured single tone RF dynamic range at both 30 kHz and 1 MHz carrier offset in the four SMuRF-RFSoC implementation’s 500 MHz wide readout bands spanning 4-4.5 GHz, 4.5-5.0 GHz, 5.0-5.5 GHz, and 5.5-6 GHz. The performance of the RFSoC data converters for many readout applications is determined by the RF dynamic range.  The measured performance exceeds the published RF performance of the legacy SMuRF system.}
   \end{figure}

In these proceedings, we focus on evaluating the readout performance for microwave SQUID multiplexers (\(\mu\)mux) coupled to low bandwidth transition edge sensors (TES), which are typically read out via flux ramp modulation at frequencies ranging from 20-80 kHz~\cite{mates12}.  To compare to the performance of the legacy SMuRF system \cite{yu2023slac}, the RF dynamic range was measured at a 30 kHz carrier offset.  To enable evaluation for applications requiring higher bandwidth, like pulse detection, the dynamic range at a 1 MHz carrier offset was also evaluated.  Figure \ref{fig:f3} shows the RF dynamic range measured for single tones over the entire 4-6 GHz SMuRF bandwidth.  The measured RF dynamic range in all four bands at a 30 kHz carrier offset fluctuates around -110 dBc/Hz, which is lower than the -100 dBc/Hz measured for the legacy SMuRF system \cite{yu2023slac}. The RF dynamic range measured at a 1 MHz carrier offset for each 500~MHz band is approximately 5~dB lower than the RF dynamic range at a 30 kHz carrier offset.

These measurements uncovered several unforeseen issues with our prototype SMuRF-RFSoC firmware port, including degradation in performance in the 4.5-5 GHz RFSoC readout band due to the unwanted image of the 4.9152 GHz ADC sampling clock against the digital numerically controlled oscillator (NCO) for down mixing at 4.75 GHz. This can be seen as a degradation of the RF dynamic range in Figure \ref{fig:f3} centered on 4.5848 GHz. We are investigating multiple approaches to solving this issue, including increasing the ADC sampling clock rate to 5 GHz.  However, as the percentage of resource utilization in the programmable logic of the RFSoC FPGA is high, increasing the ADC sample rate makes the timing closure of the firmware design challenging. We are in the process of optimizing the firmware design to meet timing requirement consistently.  We are also investigating the optimization of the RF power levels throughout the RF chain, and characterizing the RF linearity and multitone performance of the system. However for superconducting sensor applications requiring extremely low tone power levels, such as kinetic inductance detectors (KIDs), linearity requirements are not expected to be as stringent as applications like \(\mu\)mux.

\section{Readout Demonstration}

For a readout demonstration, the RFSoC system was interfaced with a Bluefors LD400 dilution refrigerator (DR) cryostat at SLAC to verify its basic functionality when connected to a \(\mu\)mux readout circuit. The DR at SLAC is instrumented with a vetted \(\mu\)mux readout chain, including a single \(\mu\)mux readout chip \cite{dober2021microwave} installed and cooled to 100 mK for this demonstration. For this basic demonstration, we ran the SMuRF-RFSoC system on a single \(\mu\)mux readout channel whose input was connected to a superconducting transition edge sensor. The sensor was kept superconducting to allow the injection and recovery of signals into the rf-SQUID's input. Because the SMuRF-RFSoC system does not yet include low frequency signal generation capabilities (required e.g. for the flux ramp and injecting signals into the TES circuit), these were provided by a legacy SMuRF system synchronized with the SMuRF-RFSoC system using an external timing system.

\begin{figure} [ht]
   \begin{center}
   \begin{tabular}{c} 
   \includegraphics[height=9.2cm]{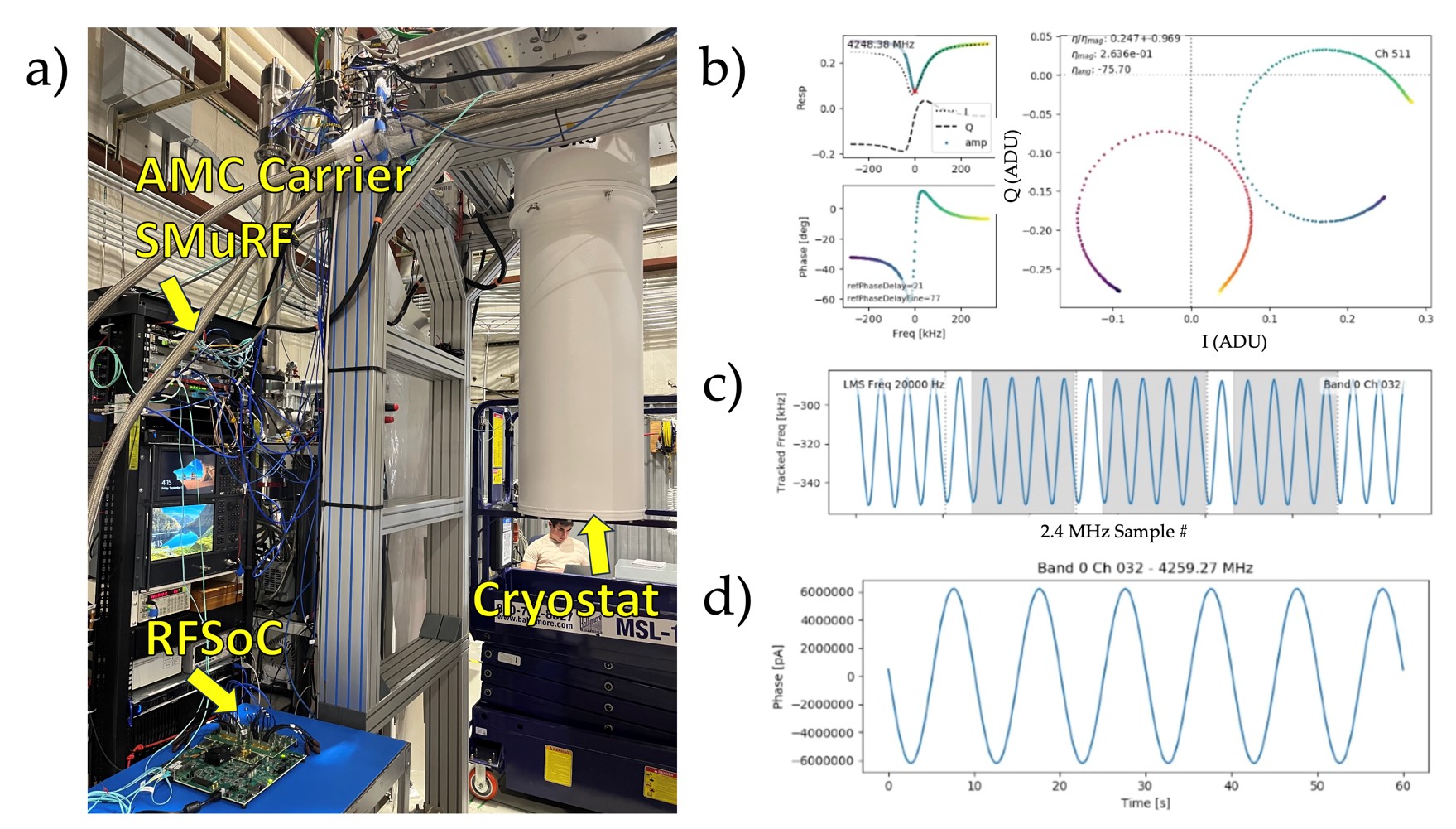}
   \end{tabular}
   \end{center}
   \caption
   { \label{fig:f4} 
a)  Test setup for first measurements using our SMuRF-RFSoC prototype to readout a superconducting sensor in the CMB dilution refrigerator DR.  b) Scan of complex transmission through a \(\mu\)mux superconducting rf-SQUID coupled resonator taken using our SMuRF-RFSoC prototype showing the expected behavior in complex transmission (amplitude and phase).  c) Tracking the modulation of the \(\mu\)mux resonator while applying a flux ramp signal taken using our RFSoC prototype, demonstrating tone tracking.  Vertical dashed lines correspond to flux ramp resets.  d) Recovered 100 mHz sine-wave signal injected into a transition-edge sensor coupled to the \(\mu\)mux channel being readout with the SMuRF-RFSoC prototype, demonstrating the successful readout of the injected signal at the expected frequency and amplitude using prototype SMuRF-RFSoC readout.}
   \end{figure}

Figure \ref{fig:f4} highlights the first successful read out of a superconducting sensor using our SMuRF-RFSoC prototype.  A small 100 mHz sine wave current signal ( \(\pm\) 6mA) injected into the rf-SQUID input is recovered with the expected frequency and amplitude.  This demonstrates all of the RF processing functionalities required for reading out TESs using the system including channelization, frequency error estimation, tone tracking, flux ramp demodulation, and tone synthesis.  We plan to build on this demonstration by systematically bench-marking the performance of the SMuRF-RFSoC prototype on superconducting transition-edge sensors coupled to resonators with frequencies spanning the full 4-6 GHz range against the legacy SMuRF system using science grade devices. This initial cryostat test aimed only to demonstrate basic functionality on a single channel, including that injected signals can be recovered at expected level, with a focus on slow signals of interest to CMB and sub-millimeter wave applications. The development and characterization of the prototype readout are now focused on demonstrating low noise performance and multichannel operation.

\section{ RFSoC Resource Utilization}  \label{Utilization}

The digital signal processing algorithms implemented in firmware for SMuRF are highly complex and challenging to fit within the programmable logic resources in RFSoC FPGA. For SMuRF-RFSoC, the logic resources in the RFSoC FPGA  still limits the number of channels that can be processed per RFSoC device, as was true for the legacy SMuRF system. As Table \ref{tab:t1} shows, although the XCZU48DR used for SMuRF-RFSoC has over 100\% more digital signal processing (DSP) slices than the XCKU15P used for SMuRF, the XCZU48DR has fewer other logic resources. While SMuRF-RFSoC eliminates the resources required for the JESD interface, which is a standardized interface between discrete data converters and FPGAs, it still requires substantial firmware resources. Due to the significant increase in DSP slices for XCZU48DR, the utilization for DSP is as low as 34\%, which is over 70\% for SMuRF system \cite{yu2023slac}. However, the reduced block random access memory (BRAM) of XCZU48DR makes its utilization rate as high as 69\%. 

The integrated ADCs in the XCZU48DR have 2.5 GHz of bandwidth, but only 500 MHz of the bandwidth is used for this implementation. If the 2 GHz of bandwidth is digitized directly and processed without segmentation, some aspects of the hardware and firmware architecture of the system could be simplified. However, the logic resources required for processing the entire band may increase dramatically, making it more challenging to meet timing requirements. 
We are investigating this and other options for optimizing the architecture of the system and its resource utilization.  In support of enabling applications with requirements beyond those of the legacy SMuRF electronics, we are exploring the parameterization of the existing SMuRF firmware blocks.

\begin{table}[ht]
\caption{Programmable logic resource comparison between the FPGA selected for SMuRF (Xilinx Kintex UltraScale+ XCKU15P) \cite{kintex} and the RFSoC used for SMuRF-RFSoC (AMD Zynq UltraScale+ RFSoC XCZU48DR). \cite{RFSoC48}}
\label{tab:t1}
\begin{center}       
\begin{tabular}{|l|l|l|} 
\hline
\rule[-1ex]{0pt}{3.5ex}  Logic Resources & XCKU15P & ZU48DR  \\
\hline
\rule[-1ex]{0pt}{3.5ex}  System Logic Cells (K) & 1,143 & 930  \\
\hline
\rule[-1ex]{0pt}{3.5ex}  DSP Slices & 1,968 & 4,272   \\
\hline
\rule[-1ex]{0pt}{3.5ex}  Memory (Mb) & 70.6 & 60.5 \\
\hline
\rule[-1ex]{0pt}{3.5ex}  I/O Pins & 668 & 347  \\
\hline 
\end{tabular}
\end{center}
\end{table}

 \begin{figure} [ht]
   \begin{center}
   \begin{tabular}{c} 
   \includegraphics[height=6cm]{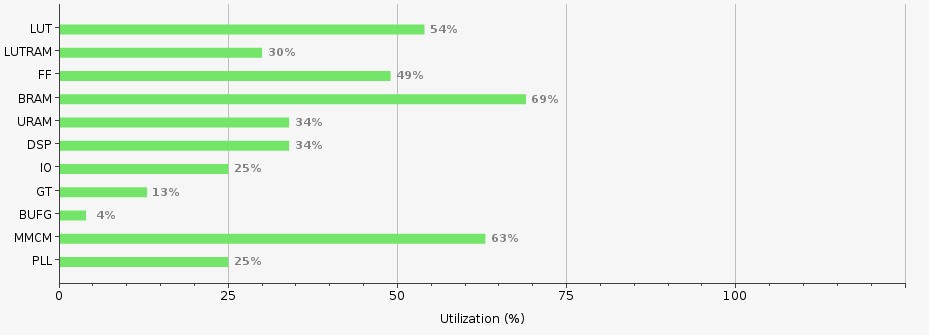}
   \end{tabular}
   \end{center}
   \caption
   { \label{fig:f5} 
The resource utilization report of firmware implementation of SMuRF-RFSoC as percentages of the total FPGA resources of an AMD Zynq UltraScale+ RFSoC XCZU48DR-2FSVG1517E device. The listed resources are as follows: look-up tables (LUT), look-up table RAM (LUTRAM), flip-flops (FF), block random access memory (BRAM), ultra RAM (URAM), digital signal processing slices (DSP), input/output interfacing (I/O),  gigabit transceiver (GT), global clock simple buffer (BUFG), mixed-mode clocking manager (MMCM), and phase-locked loops (PLL) for clocking.
}
   \end{figure} 
\section{Conclusion}

Relative to the original custom-built SMuRF system, the new SMuRF-RFSoC system we are developing has the potential for significantly reduced footprint and cost.  Among other improvements, SMuRF-RFSoC eliminates the logic resources and power required by the fast communication links required in discrete implementations between the FPGA and data converters like that of the legacy SMuRF system, as well as the need for analog RF mixers.  One benefit of this is that since the implementation of the up and down mixing is fully digital in the SMuRF-RFSoC system, it can be easily adapted to operate at different RF frequency ranges to cover other potential applications. 

Measured in loopback, the measured RF dynamic range for all four 500 MHz bands of the new SMuRF-RFSoC system is approximately 5 dB better than the that of the legacy SMuRF system.  In a separate study, we have presented a characterization of the RF linearity of the RFSoC DACs operating in higher order Nyquist zones including a characterization of their inter-modulation performance measured using two-tone tests \cite{liu2023higher}.  In this proceedings, we have demonstrated the prototype system on a single \(\mu\)mux channel, faithfully recovering an injected 100 mHz sine-wave signal.

Development is now focused on demonstrating noise performance on superconducting sensors and multitone readout. This prototype is a stepping stone towards the goal of porting the successful SMuRF system to a significantly more flexible, lower SWaP, high performance RFSoC-based readout solution for applications spanning from small scale bench-top R\&D to reading out superconducting sensor arrays instrumented in large astronomical observatories.


\acknowledgments 
 
Chao Liu, Shawn Henderson, and the development of a flexible RF readout platform for early stage R\&D in HEP and QIS were supported by the Department of Energy, Laboratory Directed Research and Development program at SLAC National Accelerator Laboratory, under contract DE-AC02-76SF00515.

\bibliography{report} 
\bibliographystyle{spiebib} 

\end{document}